\documentclass[twocolumn]{emulateapj}
\usepackage{amsmath}
\usepackage{bbding}

\newcommand{\cii}{\ensuremath{\mathrm{C}\,\scriptstyle \mathrm{II}}}
\newcommand{\oi}{\ensuremath{\mathrm{O}\,\scriptstyle \mathrm{I}}}
\newcommand{\hi}{\ensuremath{\mathrm{H}\,\scriptstyle \mathrm{I}}}
\newcommand{\heII}{\ensuremath{\mathrm{He}\,\scriptstyle \mathrm{II}}}
\newcommand{\siIV}{\ensuremath{\mathrm{Si}\,\scriptstyle \mathrm{IV}}}
\newcommand{\siIII}{\ensuremath{\mathrm{Si}\,\scriptstyle \mathrm{III}}}
\newcommand{\mgII}{\ensuremath{\mathrm{Mg}\,\scriptstyle \mathrm{II}}}
\newcommand{\p}{R$_p$/R$_*$}
\newcommand{\lya}{Lyman-$\alpha$}

\shorttitle{Limb Brightened Transits}
\shortauthors{Schlawin et al.}

\begin{document}


\title{Exo-Planetary Transits of Limb Brightened Lines; \\
Tentative Si IV Absorption by HD209458b}


\author{E. Schlawin\altaffilmark{1}}
\author{E. Agol\altaffilmark{2}}
\author{L. M. Walkowicz\altaffilmark{3}}
\author{K. Covey\altaffilmark{1,4,5}}
\author{J. P. Lloyd\altaffilmark{1}}
\altaffiltext{1}{Astronomy Department, Cornell University, Ithaca NY 14853}
\altaffiltext{2}{Astronomy Department, University of Washington, Seattle, WA 98195}
\altaffiltext{3}{Astronomy Department,University of California at Berkeley, Berkeley, CA 94720}
\altaffiltext{4}{Visiting Scholar, Department of Astronomy, Boston University, 725 Commonwealth Ave, Boston, MA 02215}
\altaffiltext{5}{Hubble Fellow}


\bibliographystyle{apj}

\begin{abstract}
Transit light curves for stellar continua have only one minimum and a  ``U'' shape. By contrast, transit curves for optically thin chromospheric emission lines can have a ``W'' shape because of stellar limb-brightening. We calculate light curves for an optically thin shell of emission and fit these models to time-resolved observations of \siIV\ absorption by the planet HD209458b. We find that the best fit \siIV\ absorption model has R$_{p,\siIV}$/R$_*$ = 0.34$^{+0.07}_{-0.12}$, similar to the Roche lobe of the planet. While the large radius is only at the limit of statistical significance, we develop formulae applicable to transits of all optically thin chromospheric emission lines.
\end{abstract}



\section{Introduction}
Since the first observation of a transiting exoplanet \citep{charbd,henryd}, knowledge of exoplanetary radii, composition and atmospheres has grown explosively. In-transit and out-of-transit spectroscopy and photometry have revealed water absorption in HD 189733b \citep{waterabs}, atmospheric emission in TrES-1 \citep{therm}, a surprising number of anomalously large planets \citep{breview}, and constraints on exoplanetary composition \citep{degen}. Accurate transit timing can also indicate the presence of additional bodies in the system through perturbations to the transiting planet's orbit \citep{agolttv,holmanttv}.

When transits are observed in stellar \lya, \cii, \siIII, \mgII\ and \oi\ emission, they show much deeper minima than for visible wavelengths, revealing escaping atmospheres extending far beyond the geometric radii\footnote{In this paper, we define the geometric radius as the radius derived from broadband visible wavelength transit depths.} of planets \citep{vidmad,benjaf7,mclay,lecav,fossati}. These light curves also constrain atmospheric conditions and mass escape from the planet's Roche lobe \citep{knutsonprop,gmunoz,linsky}.

Transit searches have been largely in the optical and near-infrared continuum, where the star is optically thick and limb darkened due to the temperature profile at the $\tau \approx$ 1 surface of the star. For limb darkened wavelengths, the flux from the transiting system is at a minimum when the planet crosses the sub-earth longitude of the star (phase = 0.0) because the stellar disk is brightest at its center.


%

By contrast, emission lines from stellar chromospheres and transition regions can be limb {\it brightened}. For optically thick emission, limb brightening occurs when the source function increases with stellar altitude. For optically thin emission, strong limb brightening occurs because the chromospheric and transition region gas has its largest column density at the edges. \citet{assef} showed that transit observations of chromospheric emission lines decrease sharply to a minimum at the first limb, increases to a local maximum mid-transit and then reverses the process as the planet exits the stellar disk. As a
consequence, such a transit curve will be ``W''-shaped.

\citet{assef} point out that limb brightening could be
useful for detecting exoplanetary transits of giant stars. Light curves of limb brightened wavelengths have
deeper minima than for both limb darkened and uniform disk
emission. The star emits over a smaller effective area--a ring instead of a disk--so the planet covers a larger amount of the
stellar flux. This is important for transits of giant stars where broadband transit depths can be below 0.01\% for Jupiter-sized planets. Also, the planet covers its host's limb for a small fraction of the transit, allowing for feasible detection of giant star transits with ground based telescopes, which suffer from systematic photometric errors over timescales longer than one night. 
The exoplanets 4UMa b, HD 122430b, HD 13189b, and HIP75458 b all have transit probabilities greater than 10\% and may be useful targets for future studies \citep{assef}.

\citet{assef} approximate the limb brightened
star as a central disk of emission surrounded by a circularly symmetric ring with $\sim$30 times the intensity. With this ring approximation, the maximum depth
of the transit is proportional to the ratio of the planet radius to stellar radius, $R_p/R_*$ if the emission from the central disk is negligible, instead of $(R_p/R_*)^2$,
as expected for a uniform disk. This is because the planet covers $\sim$2R$_p$ out of a circle of emission whose total circumference is 2$\pi R_*$.

In this paper, we present a transit light curve calculation for an optically-thin and geometrically-thin shell of emission. The calculated maximum transit depth scales as $(R_p/R_*)^{3/2}$, instead of $R_p/R_*$. In
$\mathsection$\ref{labl:chromlcurve} we calculate the expected limb
brightened light curve for an optically thin emission line. We consider optically thin emission because it shows at least 8 times the limb brightening of optically thick emission \citep{kastner}. We
fit this model light curve to \siIV\ emission from HD209458 in  $\mathsection$\ref{osiris} and discuss the implications for HD209458b's thermosphere in $\mathsection$\ref{discuss}.

\section{A Limb Brightened Curve Under The Thin-Shell Approximation} \label{labl:chromlcurve}
\label{labl:thinshell}

We make the approximation that the thickness of the chromosphere, $h$, is much smaller than the size of the planet ($R_p$) and star ($R_*$). Under this geometrically thin approximation, the total flux from the star is proportional to the surface area of the hemisphere facing the Earth
 times its thickness $h$, because the total flux for an optically thin emission line is proportional to the total number of emitting ions. Neglecting any photospheric contribution, the amount of emission that the planet blocks is then simply the amount of the stellar surface that the planet covers times $h$. In this geometrical limit, therefore, the thickness of the chromosphere, $h$, cancels out.



\begin{figure}
\begin{center}
\includegraphics[width=0.5\textwidth]{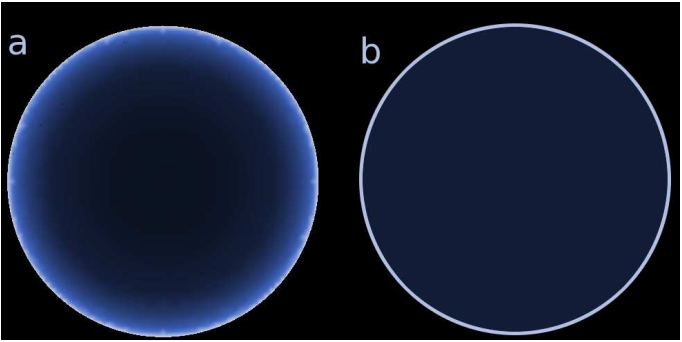}
\caption{Model Limb Brightening. (a) A model of spherically symmetric optically thin emission varies continuously from limb to center. (b) An approximate model where most of the emission is from the stellar limb surrounding a uniformly emitting circle. We employ the model shown in (a) for \siIV\ emission, whose transit light curve is given by  equation \ref{disk}.}
\label{limbmodel}
\end{center}
\end{figure}

We compute the light curve for zero thickness ($h=0$) by finding the area of the the planet's shadow and dividing this by the surface area of a hemisphere with radius $R_*$. The volume of the intersection of a cylinder with a sphere is given by \citet{lamarche} and we find the surface area of the intersection by taking a partial derivative with respect to the radius of the sphere. The result can be expressed analytically in terms of elliptic integrals.
Let $x$ be the distance, in units of stellar radii, $R_*$, from the center of the planet to the center of the star projected onto a plane perpendicular to the observer. Let $p$ be the planet/star radius ratio, \p. To calculate the light curve for a planet that does not pass through the center of the star, one can simply write $x(t)$ as $\sqrt{d^2+b^2}$, where $d$ is the distance to the closest approach point in stellar radii $R_*$ and $b$ is the impact parameter in stellar radii $R_*$. 

The transit depth, $\delta (x)$ is a piecewise function with three different regimes: (1) when the planet is fully contained in the stellar disk (2) when the planet is at egress/ingress and (3) when the planet is beyond the stellar disk. We also include the case that the planet's absorption profile is larger than the star. In this case, the same formulae apply--see Equation \ref{defs}. A small figure for each regime is included in the equations, where an open circle represents the star and a filled circle represents the planet.

\begin{eqnarray}\label{disk}
\delta(x)&=& \Theta (p-x)+\frac{a_0}{2 \pi \sqrt{xp}} \times \\
&&\left[\frac{x+p}{x-p} \Pi(n,m) - 4 x p a_1 E(m) - a_2 K(m) \right] \nonumber
\end{eqnarray}


\begin{align} \label{defs}
&& \ooalign{$\bigcirc$\cr\kern+3pt$\bullet$} && \ooalign{$\bigcirc$\cr\kern+7.5pt$\bullet$}  && \bigcirc ~ \bullet \nonumber \\
p < 1&&0 < x < 1-p & & 1 - p < x < 1+p  & & 1+p < x \nonumber \\
\hline
&&  && \ooalign{$\circ$\cr\kern+2pt\CircleSolid}  && \ooalign{\CircleSolid} ~ \mathrm{or} ~ \circ ~ \ooalign{\CircleSolid} \nonumber \\
p > 1 && &&  p - 1 < x < 1 + p && 0 < x < p - 1 \nonumber \\
&& && && \mathrm{or} ~ x > p + 1 \nonumber \\
\hline
a_0 && \sqrt{m} && 1 && 0 \nonumber \\
a_1 && \frac{1}{m} && 1 && \nonumber \\
a_2 && x^2-p^2 && 1-2p(x+p) && \nonumber \\
m && 4xp\over 1-(x-p)^2 && \frac{1-(x-p)^2}{4xp} && \nonumber \\ 
n && -4xp/(x-p)^2 && 1-(x-p)^{-2} &&
\end{align}


where $K(m)$, $E(m)$, and $\Pi(n,m)$ are the complete Legendre elliptic 
integrals of the first, second, and third kinds, and $\Theta(x)$ is
the Heaviside step function. For the elliptic integrals, we use the conventions of \citet{handbk} where $\Pi(n,m) = \Pi(n;K(m)|m)$ for the third elliptic integral. \footnote{An IDL procedure for the transit depth $\delta (x)$ is located at \texttt{http://www.astro.washington.edu/agol/}.}

These formulae are difficult to evaluate numerically at 
$x=0$, $x=p$ and $x=1 \pm p$ due to the formal divergence of different terms in equations \ref{disk} and \ref{defs}; the divergences cancel out analytically, but routines that
evaluate the elliptic integrals diverge.  However, these locations
are a set of measure zero, and thus are tractable
when modeling data. 

\subsection{Analytical Transit Depth Estimation}

\begin{figure}
\begin{center}
\includegraphics[width=0.5 \textwidth]{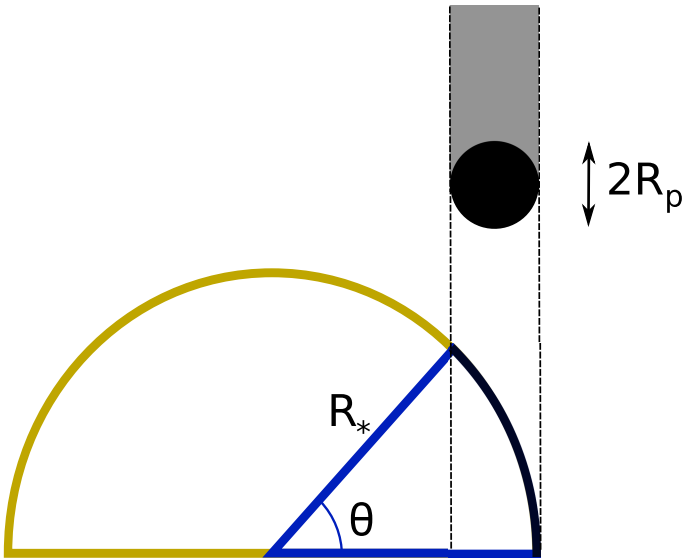}
\caption{Edge-on view of area of stellar emission and the amount blocked by  a planet. This blocked surface area is the same as the area of a shadow cast by a sphere onto a hemisphere.}
\label{fig01}
\end{center}
\end{figure}

The transit is deepest slightly before second contact, so we can estimate the maximum transit depth as follows from
Figure \ref{fig01} by comparing the total emitting area of the star to the total stellar surface area blocked by the planet. The blocked surface is the same as the shadow produced by a sphere of radius $R_p$ onto a hemisphere of radius $R_*$. When the planet occults the edge of the star 
(second contact), as shown from an edge-on viewpoint in this figure,
then the length of the arc of the long axis of the shadow is
$R_*\theta$.  The diameter of the planet is 
$2 R_p \approx R_*(1-\cos{\theta}) \approx \frac{1}{2} R_* \theta^2$,
where the latter approximation is valid for $\theta \ll 1$.
 We can approximate
the shadow as an ellipse with a semi-minor axis of $R_p$
and a semi-major axis of $\frac{1}{2} R_*\theta$, so the
area of the shadow is $A_t = \pi \sqrt{R_pR_*} R_p$.  Thus,
the maximum depth of transit is given by 
\begin{equation} \label{depth_anal}
\delta_{\mathrm{max}} \approx {A_t \over 2 \pi R_*^2} = \frac{1}{2} \left({R_p \over R_*}\right)^{3/2}
\end{equation}
which is accurate to within 5\% for \p $<$ 0.23. Note that this is a different scaling for maximum transit depth than that given in \citet{assef} who assume that most of the stellar emission is from a thin ring. For \p\ = $1-\sqrt{3} /2$, half of the stellar emission is within $R_p$ of the stellar radius $R_*$, so for \p\ $\gtrsim 0.13$, the ring approximation is valid, but the scaling of transit depth assumes negligible limb curvature at the scale of the planet (\p\ $\ll$ 1).


The remarkable consequence of Equation \ref{depth_anal} is that the depth of a chromospheric
transit does not decline as much with the radius of the planet as 
a transit of a uniform disk.  A chromospheric transit has a maximum
depth that is $\approx \frac{1}{2} \left({R_* \over R_p}\right)^{1/2}$
times deeper than the maximum transit depth of a uniform disk;
thus smaller planets have an advantage to be observed at
chromospheric wavelengths, as emphasized by \citet{assef}. It should also be noted that at mid-transit, the Double-U curve has a smaller transit depth than for a uniform disk, because the planet covers only $\pi R_p^2$ out of a hemisphere of area $2 \pi R_*^2$.


Figure 3 (solid curve) shows the transit light curve, 1-$\delta (x)$, for a planet that has \p =0.08. The estimate given by equation \ref{depth_anal} and the more detailed equation \ref{disk} agree well, predicting maximum chromospheric depths to be 1.8 times
deeper than for uniform disk brightness. (For a uniform disk emission $\delta_{\mathrm{max}} $=$\pi R_p^2 /(\pi R_*^2)$). We also include a light curve for the thin circle of emission shown in Figure \ref{limbmodel} b for comparison.




\begin{figure}
\begin{center}
\includegraphics[width=0.5 \textwidth]{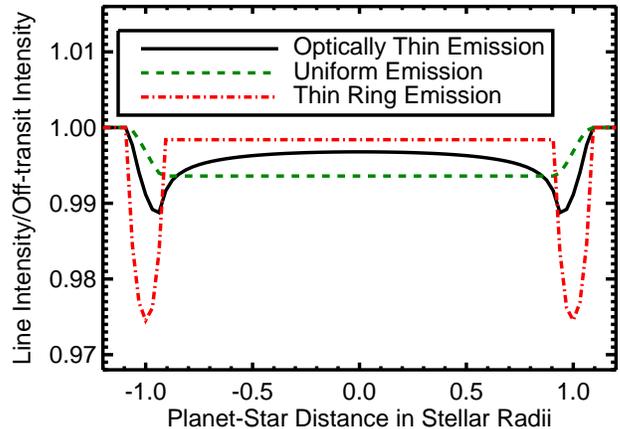}
\caption{Transit light curve for \p $=0.08$, using a family of 3 different models. The solid line is for a transit of an optically thin shell using equation \ref{disk} where the emission looks like Figure \ref{disk} a. The ``Uniform Emission'' \citep[dashed green line]{magol} is one for which the emission is assumed to be constant across the stellar disk. The ``Thin Ring Emission'' model (dot-dash red line) is a model where the emission is assumed to be mostly from a thin ring as pictured in Figure \ref{limbmodel} b. Note that the ``Thin Ring'' and ``Optically Thin Emission'' models' minima are deeper than the uniform emission model and that these minima occur near the stellar limbs. }
\end{center}
\label{fig02}
\end{figure}

\begin{figure}
\begin{center}
\includegraphics[width=0.5\textwidth]{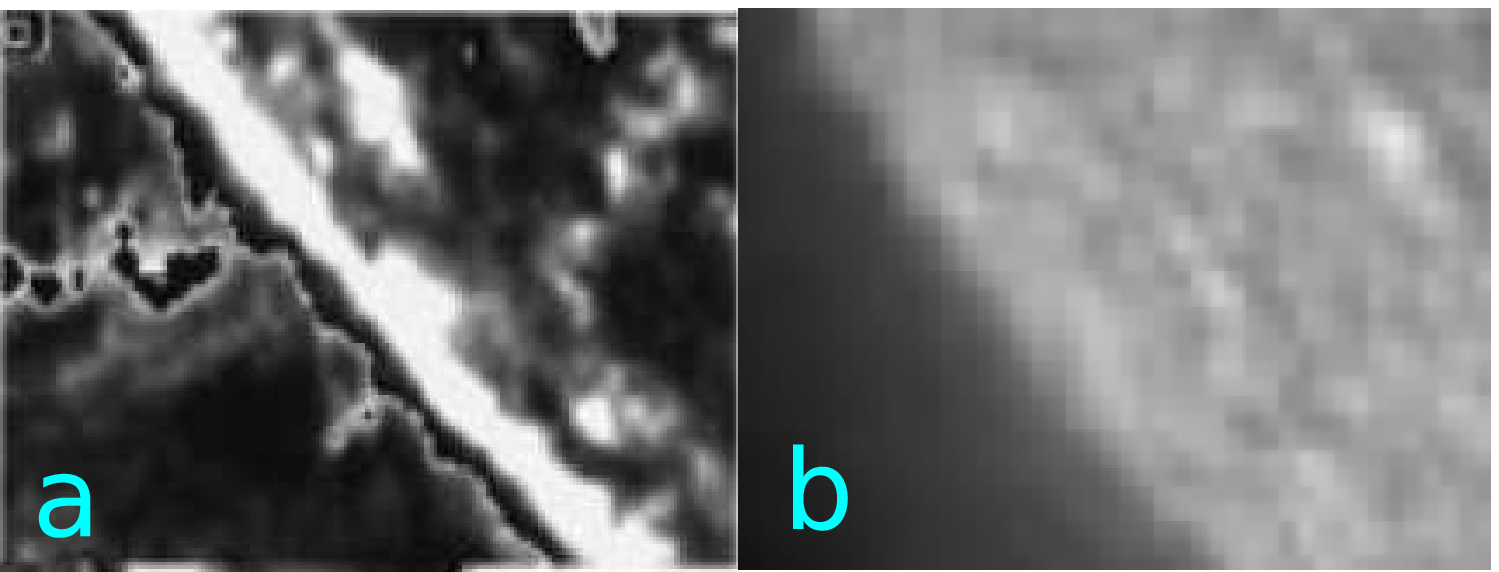}
\caption{(a) \siIV\ emission from the Solar limb as observed by SUMER (Solar Ultraviolet Measurement of Emitted Radiation) \citep{wiik} is strongly limb-brightened, indicating that a transit of this emission line should have a Double-U light curve. (b) \heII\ 304 \AA\ image of the solar limb, taken by the Extreme-ultraviolet Imaging Telescope \citep{feldman} is, by contrast, not limb brightened because it is optically thick in the chromosphere and thus the additional column density at the limb does not contribute any more flux than the central disk.}
\label{limbs}
\end{center}
\end{figure}

\section{\siIV\ Absorption by HD 209458b} \label{osiris}

\citet{viddisc,vidmad} observed the exoplanet host HD 209458 with the Hubble Space Telescope Imaging Spectrograph (STIS) instrument and found
an extended hydrogen, oxygen and carbon atmosphere around the planet by fitting the light curves
to the \hi, \oi\ and \cii\ lines during a transit. \citet{vidmad} found no \siIV\ absorption when fitting the light curve to a model of a spherical planet occulting a {\it uniform} stellar disk.

We fit their \siIV\ 1394 \AA\ data (\citet{vidmad}, Figure 3) with a Double-U model given by equation \ref{disk}. This model is appropriate for the \siIV\ emission, as evident in the Solar image in Figure \ref{limbs} (a) where strong limb-brightening is apparent. \citet{france10} found that the \siIV\ 1394 \AA / \siIV\ 1403 \AA\ ratio is 2:1 within the errors, indicating that the \siIV\ 1394 \AA\ emission is optically thin \citep{bloom02,christian06}. We investigated the transit of the \siIV\ line because it the strongest optically thin line in HD209458b's STIS spectrum. While other limb brightened emission lines do exist, we focus on the optically thin ones because they should be the most limb brightened.

The model has two free parameters: the planet/star radius ratio \p\ and an overall constant that sets the off-transit flux. The second parameter is necessary since the off-transit \siIV\ 1394 \AA\ STIS flux is poorly constrained. The impact parameter 
 is fixed with a value of $ b= 0.50 R_*$, using an inclination of 86.7$^\circ$ and a semi-major axis $a=8.76 R_*$ \citep{torres}.

Figure 5
 shows the light curve for the \citet{vidmad} data and the best fit model. In addition to the best fit model, we show two more limb brightened models where \p\ is a fixed parameter for comparison. These have \p\ equal to 0 and 0.12, representing no \siIV\ absorption and the geometric planet radius \citep{knutsonprop}, respectively.

Using a Levenberg-Marquardt fitting algorithm \citep{mpfit} and the points at which the $\chi^2 - \chi_{min}^2$ = 1 to calculate uncertainties, we find that \p\ = 0.34$^{+0.07}_{-0.12}$, close to the size of the Roche Lobe. The Double-U model has a total $\chi^2$ of 10.4 with 12 data points and two parameters, which is 3.1 less than fitting the data to
a constant flux line (representing no detection) with one parameter. Since these models are nested, we can use the maximum likelihood ratio test. The P-value for the difference in $\chi^2$ is 0.05, so the data favor the Double-U model with 95\% confidence, assuming normally distributed data. The same test favors the \p\ =0.34 model over the \p\ = 0.12 model with 91\% confidence.


\begin{figure}[!ht]
\begin{center}
\includegraphics[width=0.5 \textwidth]{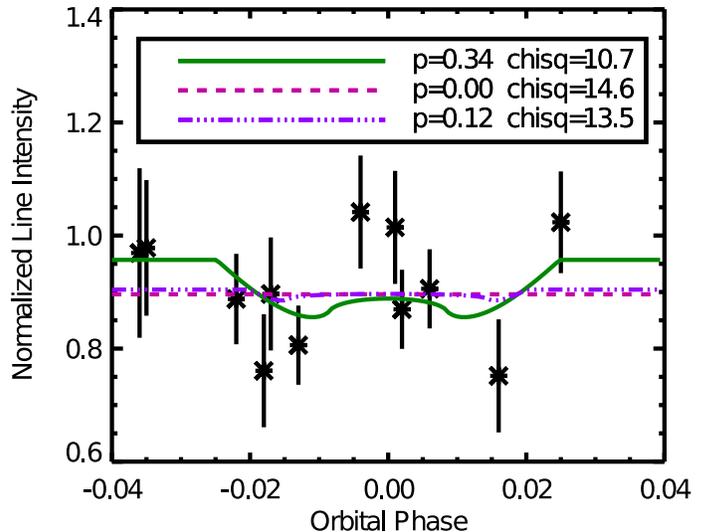}
\caption{The light curve of the limb-brightened transition region line \siIV\ during the transit of HD209458b, as taken by \citet{vidmad} for a wavelength range of [1391 \AA,1397 \AA]. Fluxes are normalized so that the weighted average of the points when the planet is not occulting the star is 1.0. The solid line is a best fit model, using equation \ref{disk} where the radius is a free parameter. For reference, two different best-fit models are shown where the planet size is a fixed parameter: the dashed curve is a horizontal line, corresponding to R$_p$/R$_*$ = 0 or a non-detection of \siIV\ absorption. The dash triple-dotted line is a model with the geometric radius, R$_p$/R$_*$ = 0.12. The best-fit radius, R$_p$/R$_*$=0.34 (normalized $\chi^2$ = 1.07), indicates \siIV\ absorption from a Roche lobe-sized cloud.}
\end{center}
\label{lightc}
\end{figure}

\section{Conclusion}

As indicated by the Solar image of \siIV\ in Figure \ref{limbs}, a limb brightened model should be used for \siIV\ emission. The Double-U model for the \siIV\ transit fits the data better than a non-detection, even when accounting for the additional parameters in the model. The best fit absorption radius \p=0.34$^{+0.07}_{-0.12}$, if the absorption is entirely optically thick. This radius favors planetary atmosphere models with mass flow beyond the planet's Roche lobe.

The best time to observe transits in optically thin stellar emission lines is when a planet crosses its host's limb and not at a phase of 0, since the Double-U curve is deepest at the stellar limb. Observations at the limb have a depth of $\sim$0.5 (\p )$^{3/2}$ whereas at the center of the star they have a transit depth of  $\sim$0.5 (\p)$^2$, only half the depth of a uniform brightness transit.

\subsection{Discussion} \label{discuss}

The \siIV\ transit depth is comparable to the \oi\ transit depth calculated by \citet{vidmad} and larger than the \hi\ transit depth calculated by \citet{benjaf7}. As \citet{kosk} point out, a hard sphere, which we assume in equation \ref{disk}, is a poor approximation to planetary atmospheric absorption. The \siIV\ absorption profile, instead of having a sharp transition from opaque to transparent, should change smoothly from optically thick to optically thin absorption in a more accurate model. In order to explain the large transit depth, this smooth model would have to have a radius extending beyond the planet's Roche lobe to explain the observed transit depth. R$_{\mathrm{Roche}}$/R$_*$ varies from 0.35 to 0.48 \citep{ben10} .

The best-fit transit depth supports models with high concentrations of metallic ions in the atmosphere because the radius of \siIV\ absorption is as large as for \hi\ absorption. \citet{kosk} point out that if the metallicity is high, the temperature must also be elevated. The $\sim$10,000 K temperature suggested by \citet{gmunoz}, \citet{mclay}, and \citet{kosk} for the thermosphere may not explain the large abundances of Si$^{3+}$ needed for the observed absorption. The ionization energy, $\Delta E/k$ for Si$^{2+} \rightarrow$ Si$^{3+}$ = $3.9 \times 10^5$K, suggests that the Si$^{3+}$ may be produced in a shock between the stellar and planetary winds.


\citet{linsky}
find no significant \siIV\ absorption by HD209458b.
Our best fit to the STIS data predicts the average transit depth to be 9$^{+4}_{-5}$\% for the same phases as their observations, but \citet{linsky}, with the Cosmic Origins Spectrograph (COS), found 0.2 $\pm$ 1.4\%. These results disagree at the $\sim$1.7$\sigma$ level and variability in the planet's atmosphere may account for the discrepancy. \citet{linsky} did observe some weak \siIV\ absorption features found at +20 and +40km/s in their spectrum, which indicates that some Si$^{3+}$ ions remain in the planet's atmosphere or winds.

\citet{linsky} suggest that the amount of another Silicon ion, Si$^{2+}$, may vary appreciably over short timescales because of changes in stellar wind speed, planetary mass-loss rate or temperature fluctuations. Conversely to \siIV , for which 2003 STIS observations indicate strong absorption and 2009 COS observations indicate weak absorption, the \siIII\ absorption is seen strongly in absorption in the 2009 COS data and weakly in the 2003 STIS data.

Additional observations will help confirm or refute the detection of Si$^{3+}$ in the thermosphere of HD209458b. The Cosmic Origins Spectrograph (COS) is an ideal instrument with 2 to 10 times the sensitivity of previous ultraviolet spectrographs \citep{cos}. COS was used by \citet{linsky}, but only when HD209458b was close to a phase of 0.0, 0.25, 0.5 and 0.75 and once when the planet was at the stellar limb. Additional observation at the host's limb would be optimal for \siIV\ and other optically thin emission lines. With enough signal to noise, the light curve may reveal asymmetries in the transit having to do with an asymmetric spatial distribution of the UV-absorbing cloud. The advantage of the limb brightened emission lines is that they come from a smaller spatial area of the star and therefore probe the spatial distribution of the planetary atmosphere better than optically thick emission. Accurate time resolved transit data may also reveal differences in {\it thermal} properties of the leading and trailing sides of the planet as predicted by \citet{fortney}.


E. Schlawin was supported by the NASA Space Grant Fellowship. E. Agol was supported in part by the NSF under Grant No.
PHY05-51164 during a visit to the Kavli Institute for
Theoretical Physics, and by NSF CAREER Grant No. 0645416. Support for Kevin Covey was provided by NASA
through Hubble Fellowship grant \#HST-HF-51253.01-A awarded by the
Space Telescope Science Institute, which is operated by the
Association of Universities for Research in Astronomy, Inc., for NASA,
under contract NAS 5-26555. Lucianne Walkowicz is grateful for the support of the Kepler Fellowship for the Study of Planet-Bearing Stars.
\\


\end{document}